\documentclass[aps,prb,twocolumn,superscriptaddress]{revtex4-1} %linenumbers
\usepackage{graphicx}
\usepackage{hyperref}
\usepackage{color}
\usepackage{todonotes}
\begin{document}

\title{Itinerant effects and enhanced magnetic interactions in Bi-based multilayer cuprates}

\author{M. P. M. Dean}
\email{mdean@bnl.gov}
\affiliation{Department of Condensed Matter Physics and Materials Science, Brookhaven National Laboratory, Upton, New York 11973, USA}
\author{A. J. A. James}
\affiliation{Department of Condensed Matter Physics and Materials Science, Brookhaven National Laboratory, Upton, New York 11973, USA}
\affiliation{London Centre for Nanotechnology, University College London, London WC1E 6BT, United Kingdom}

\author{A. C. Walters} 
\affiliation{Diamond Light Source Ltd., Harwell Science and Innovation Campus, Chilton, Didcot, Oxfordshire, OX11 0DE, UK}
\author{V. Bisogni} 
\affiliation{Photon Sciences Directorate, Brookhaven National Laboratory, Upton, New York 11973, USA}
\affiliation{Swiss Light Source, Paul Scherrer Institut, CH-5232 Villigen PSI, Switzerland}

\author{I. Jarrige} 
\affiliation{Photon Sciences Directorate, Brookhaven National Laboratory, Upton, New York 11973, USA}

\author{M. H\"{u}cker}
\affiliation{Department of Condensed Matter Physics and Materials Science, Brookhaven National Laboratory, Upton, New York 11973, USA}

\author{ E. Giannini}
\affiliation{D\'{e}partement de Physique de la Mati\`{e}re Condens\'{e}e, Universit\'{e} de Gen\`{e}ve, CH-1211 Gen\`{e}ve 4, Switzerland}
\author{M. Fujita} 
\affiliation{Institute for Materials Research, Tohoku University, Katahira, Sendai 980-8577, Japan}

\author{J. Pelliciari}
\author{Y. Huang}
\affiliation{Swiss Light Source, Paul Scherrer Institut, CH-5232 Villigen PSI, Switzerland}
\author{R. M. Konik}
\affiliation{Department of Condensed Matter Physics and Materials Science, Brookhaven National Laboratory, Upton, New York 11973, USA}
\author{T. Schmitt}
\affiliation{Swiss Light Source, Paul Scherrer Institut, CH-5232 Villigen PSI, Switzerland}
\author{J. P. Hill}
\email{Hill@bnl.gov}
\affiliation{Department of Condensed Matter Physics and Materials Science, Brookhaven National Laboratory, Upton, New York 11973, USA}

% user macros
\def\mathbi#1{\ensuremath{\textbf{\em #1}}}
\def\Q{\ensuremath{\mathbi{Q}}}
\newcommand{\angstrom}{\mbox{\normalfont\AA}}

\date{\today}

% shorter version
\begin{abstract}
The cuprate high temperature superconductors exhibit a pronounced trend in which the superconducting transition temperature, $T_{\rm c}$, increases with the number of CuO$_2$ planes, $n$, in the crystal structure. We compare the magnetic excitation spectrum of Bi$_{2+x}$Sr$_{2-x}$CuO$_{6+\delta}$ (Bi-2201) and  Bi$_2$Sr$_2$Ca$_2$Cu$_3$O$_{10 + \delta}$ (Bi-2223), with $n=1$ and $n=3$ respectively, using Cu $L_3$-edge resonant inelastic x-ray scattering (RIXS). Near the anti-nodal zone boundary we find the  paramagnon energy in Bi-2223 is substantially higher than that in Bi-2201, indicating that multilayer cuprates host stronger effective magnetic exchange interactions, providing a possible explanation for the $T_{\rm c}$ vs.\ $n$ scaling. In contrast, the nodal direction exhibits very strongly damped, almost non-dispersive excitations. We argue that this implies that the magnetism in the doped cuprates is partially itinerant in nature.
\end{abstract}

%\keywords{}

\maketitle
%\section{Introduction}
%Since the discovery of high-$T_{\rm c}$ superconductivity in the cuprates many different variety of cuprates have been synthesized. 
Despite over 25 years of research we still have very few reliable strategies for increasing the superconducting transition temperature, $T_{\rm c}$, of the cuprates. One well established and widely applicable method is to vary the number of neighboring CuO$_2$ planes, $n$, in the crystal structure. $T_{\rm c}$ typically increases from $n=1$ to $n=3$, before dropping off for $n \geq 4$ where high quality single crystal samples have proved challenging to synthesize.\cite{Iyo2007} In Bi-based cuprates the maximum $T_c$ is 34~K for $n=1$ and 110~K for $n=3$ with Hg-, and Tl-based cuprates showing similar trends.\cite{Feng2002,Tarascon1988, DiStasio1990, Scott1994} This has lead to numerous competing explanations. Some researchers suggest that tunneling of Cooper pairs between the layers can enhance $T_{\rm c}$.\cite{Kivelson2002, Chakravarty2004, Nishiguchi2013} Alternatively, the outer CuO$_2$ layers have been proposed to protect the inner layers from the disorder present in the cuprate spacer layers.\cite{Eisaki2004, Fujita2005, Alloul2009} The strength of the next nearest neighbor (and other higher order) hopping parameters have also been implicated as a mechanism for increasing $T_{\rm c}$.\cite{Pavarini2001, Sakakibara2014} Other work suggests that changes in $T_{\rm c}$ are dependent on the energy of the $p_z$ orbitals from the apical oxygens.\cite{Ohta1991} Finally,  it may be that screening of plasmon-like intralayer modes leads to an increase in $T_{\rm c}$ with $n$.\cite{Leggett1999}

\begin{figure}
\includegraphics{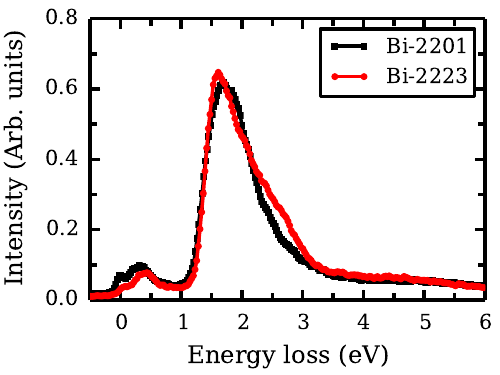} %
\caption{(Color online) RIXS Spectra for Bi-2201 and Bi-2223 taken at $\mathbi{Q}=(0.39,0)$ . Magnetic, orbital and charge-transfer excitations are seen in energy ranges around $0-400$~meV, $1-3$~eV and $2-8$ eV respectively (high energy scales $>6$~eV are not shown). }
\label{Fig1}
\end{figure}
More detailed spectroscopic measurements of the properties of multilayer cuprates are crucial in distinguishing between these different proposals. High quality angle-resolved photoemission (ARPES) measurements of trilayer Bi-cuprates have recently been obtained.\cite{Ideta2010} These data argue against proposals based on changes in the next-nearest neighbor hopping parameters.\cite{Pavarini2001} Another ARPES study reported that the magnitude of the ``kink'' in the electronic structure increases from $n=1$ to $n=3$ suggesting that multilayer cuprates host a stronger electron-boson coupling.\cite{Matsui2003} Unfortunately, to date there is very limited information about the magnetic interactions in the multilayer cuprates. This is because these systems are usually very difficult to synthesize in the large single crystal form required for inelastic neutron scattering (INS). Relatively recently, Cu $L_3$-edge resonant inelastic x-ray scattering (RIXS) has emerged as an alternative probe of the magnetic excitations in the cuprates.\cite{Dean2014} Here we use RIXS to probe and compare the magnetic excitations in single layer (Bi-2201) and trilayer (Bi-2223) bismuth-based cuprates. We report two main conclusions. First, we find that the paramagnon energy along the anti-nodal (Cu-O bond) direction is substantially higher in Bi-2223 compared to Bi-2201. This is consistent with the hypothesis that, everything else being equal, $T_{\rm c}$ scales with the strength of the effective Cu-Cu nearest neighbor magnetic exchange interaction, $J$.\cite{Pines2013} Such an appealingly simple scenario must be taken into account against competing proposals for the $T_c$ vs.\ $n$ scaling relationship. Second, we find that the nodal direction in the Brillouin zone hosts very strongly damped, almost non-dispersive excitations in both Bi-2201 and Bi-2223, in stark contrast to the anti-nodal direction. This phenomenology can restrict models of magnetism in the cuprates and we show that it is captured by itinerant calculations of the dynamical magnetic susceptibility.\cite{Brinckmann2001} This implies that the nature of the magnetism in the doped cuprates is at least partially itinerant. 

%\section{Methods}
Bi$_{2+x}$Sr$_{2-x}$CuO$_{6+\delta}$ (Bi-2201) $x=0.20$ single crystals were grown using the floating-zone method. The hole concentration for $x=0.20$ was determined to be $p=0.12(1)$ using Hall effect measurements.\cite{Enoki2013} This particular Bi-2201 system was chosen because it has an especially low  $T_{\rm c} \approx 1$~K due to its narrow $T_{\rm c}-p$ dome which peaks at 9~K.\cite{Enoki2013, Kudo2009} Bi$_2$Sr$_2$Ca$_2$Cu$_3$O$_{10 + \delta}$ (Bi-2223) single crystals were also prepared using the floating zone method.\cite{Giannini2004} An onset $T_{\rm c}= 109$~K with a width of $\sim4$~K, was measured by SQUID magnetometry putting it in the slightly underdoped region,\cite{Giannini2004} similar to the Bi-2201 sample, though the precise doping level of Bi-2223 is difficult to define, because the different CuO$_2$ layers have different doping levels.\cite{Ideta2010} The samples were cleaved to reveal a fresh surface immediately before introducing them into vacuum. RIXS spectra were measured with the SAXES spectrometer located at the ADRESS beamline  of the Swiss Light Source at the Paul Scherrer Institut.\cite{Ghiringhelli2006,Strocov2010} The x-ray energy was set at the peak in the Cu $L_3$-edge total fluorescence yield x-ray absorption spectrum at approximately 931~eV. The experimental resolution and the elastic energy of the RIXS spectra were determined by measuring the diffuse elastically scattered x-rays from disordered graphite, giving a resolution of $\sim$140~meV Gaussian full-width-half-maximum (FWHM). The elastic energy was further confirmed by observing that the energy of the $dd$ excitations remained independent of $\mathbi{Q}$, as expected for localized excitations.  We present spectra normalized to the integrated intensity of the $dd$-excitations, as was done in previous studies.\cite{LeTacon2011, DeanBSCCO2013, DeanLSCO2013} The scattering vector, \mathbi{Q}, is denoted using the pseudo-tetragonal unit cell $a \approx b \approx 3.8$~\AA{}. X-rays were scattered through a fixed angle of $130^{\circ}$ and the sample was rotated to change $Q_{\parallel}$, the projection of \mathbi{Q} into the CuO$_2$ planes. Here high $Q_{\parallel}$ corresponds to close-to-grazing exit geometry.  The x-ray polarization was parallel ($\pi$)  to the horizontal scattering plane and all data were taken at low temperature, $T \approx 15$~K.

RIXS spectra showing the $dd$-orbital transitions in Bi-2201 and Bi-2223 are plotted in Fig.~\ref{Fig1}. The Bi-2223 sample shows additional spectral weight, compared to Bi-2201, around $\sim 2.7$~eV relative to the main peak at $1.6$~eV. The inner CuO$_2$ planes in Bi-2223 lack apical oxygens, meaning that the Cu atoms in these layers are in the strongly tetragonal crystal field limit, pushing the $z^2$ orbital transition  to higher energies. The spectra in Fig.~\ref{Fig1} are consistent with the additional spectral weight in Bi-2223 coming from excitations into the high-energy orbitals in the inner CuO$_2$ planes. Such a scenario should be tested by quantum chemical calculations  and detailed measurements of the RIXS polarization dependence in future experimental studies.\cite{Hozoi2011}

\begin{figure}
\includegraphics{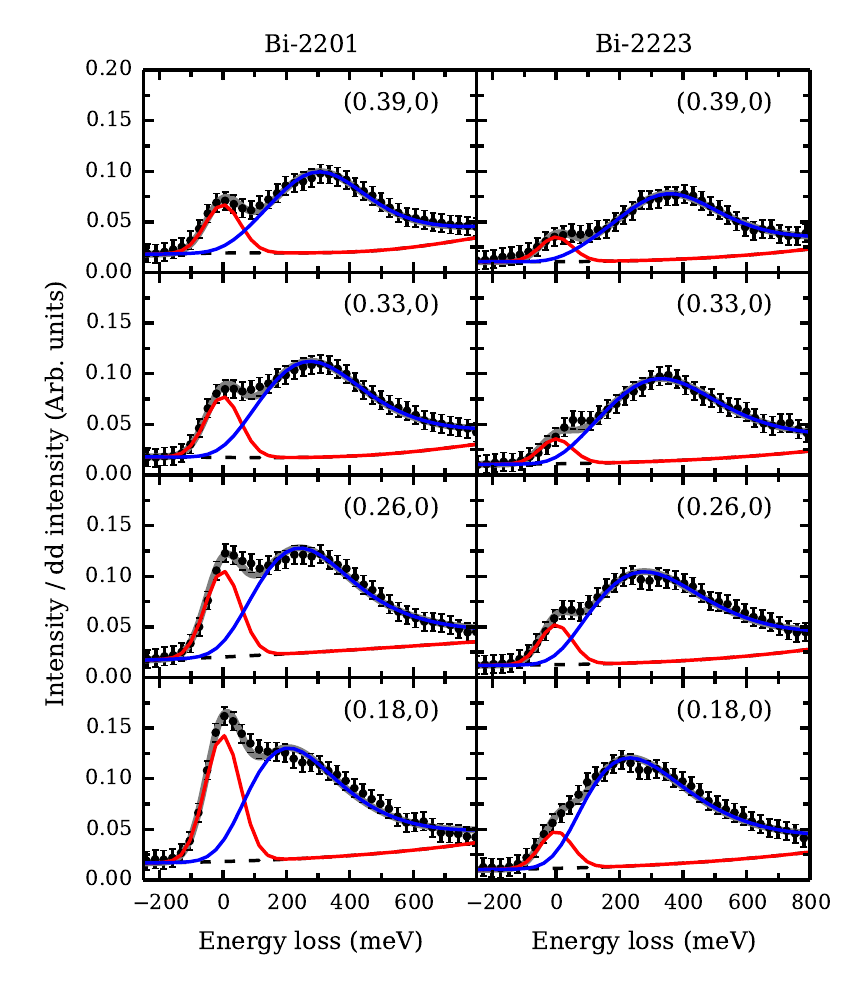} %
\caption{(Color online) Low energy RIXS spectra for Bi-2201 (left) and Bi-2223 (right) along $(\zeta, 0)$ showing the dispersion of the paramagnon excitation. Black points show the data and the solid grey line represents the results of the fitting, which is the sum of a Gaussian elastic scattering contribution (red), an anti-symmetrized Lorentzian capturing the magnetic scattering (blue), and the smooth background (dashed black line).}
\label{Fig2}
\end{figure}

The dispersion of the low energy RIXS spectra along the antinodal $(\zeta,0)$ direction, where $\zeta$ is a reciprocal space coordinate, is shown in Fig.~\ref{Fig2}. A broad, dispersive paramagnon mode is observed -- much like the spectra for other doped cuprates and consistent with Hubbard model calculations, which also predict the presence of paramagnons in the doped cuprates.\cite{Dean2014, Jia2014} The line shape was modeled using a resolution-limited Gaussian function to account for elastic scattering and an anti-symmeterized Lorentizian multiplied by the Bose-Einstein distribution and convolved with the instrumental resolution to account for the paramagnon excitation. A smooth background, obtained by interpolating between the energy gain region of the spectrum below about -100~meV and the $dd$ range of the spectrum above about 800~meV, was used to account for the tail of the $dd$-excitations. A similar procedure was used in Refs.~\onlinecite{LeTacon2011, DeanLSCO2013, DeanLBCO2013}.

\begin{figure}
\includegraphics{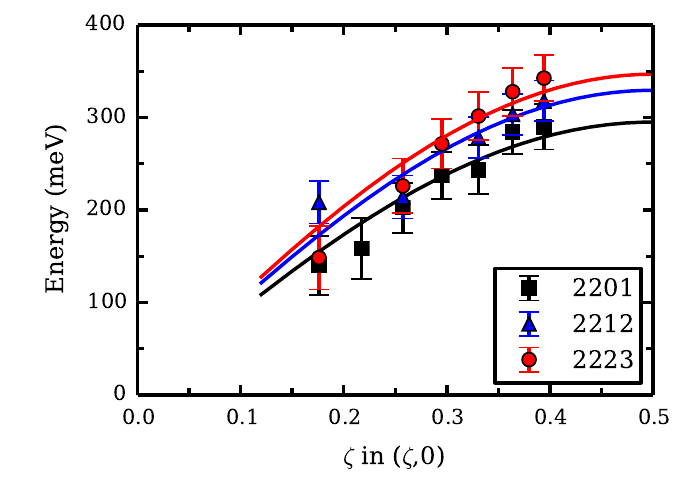} %
\caption{(Color online) The dispersion of the energy of the paramagnon mode in Bi-2201 and Bi-2223 along $(\zeta, 0)$, as obtained by the fitting analysis of the spectra in Fig.~\ref{Fig2}. The points for Bi-2212 are obtained by fitting the data in  Ref.~\onlinecite{DeanBSCCO2013} in the same way. The lines are sinusoidal fits to the Bi-2201, Bi-2212 and Bi-2223 data, which serve as guides to the eyes. Errorbars are determined by summing the estimated uncertainly in the elastic energy and the statistical error from the least-square fitting routines in quadrature. }
\label{Fig3}
\end{figure}

Figure \ref{Fig3} plots the paramagnon energy as a function of $\zeta$ along $(\zeta,0)$ as determined by this fitting procedure. The energy scale of the paramagnon in Bi-2223 is consistently higher than that in Bi-2201 and the values for Bi-2212 tend to fall in between these two limits. Assuming a spin-wave-like phenomenological functional form where the energy, $E \propto\sin(\mathbi{Q} \cdot \mathbi{a})$ we plot guides to the eye on Fig.~\ref{Fig3}. Extrapolating these lines to the $(0.5,0)$ zone boundary gives paramagnon energy scales of 295(21)~meV,  329(14)~meV and 347(14)~meV for Bi-2201, Bi-2212 and Bi-2223 respectively. We take the magnitude of the zone boundary paramagnon energy as a measure of the effective nearest-neighbor magnetic exchange interaction, $J$. Single layer doped cuprates such as Yttrium-based systems \cite{LeTacon2011} and La$_{2-x}$Sr$_x$CuO$_4$ [Ref.~\onlinecite{DeanLSCO2013}] also show energy scales similar to Bi-2201 as does single layer Tl$_2$Ba$_2$CuO$_4$,\cite{LeTacon2013} although Tl$_2$Ba$_2$CuO$_4$ has only been studied in the overdoped region of the phase diagram. Several previous studies have shown that the doping dependence of the paramagnon energy in the cuprates is rather weak,\cite{LeTacon2011, DeanBSCCO2013, DeanLSCO2013, LeTacon2013, Dean2014} which implies that the small difference in doping between the Bi-2201 and Bi-2223 samples is not a determining factor for the observed difference in the paramagnon energies. Similarly, independent work has shown that with the current energy resolution no significant change in the paramagnon energy is seen through the superconducting transition.\footnote{Private communication: Thorsten Schmitt.} 

% WIDTHS COMMENT
The widths of the paramagnon excitations are approximately 250~meV (Lorentzian half width at half maximum) and roughly independent of $\mathbi{Q}$. Within errors we do not find significant differences between Bi-2201 and Bi-2223. In multilayer cuprates such as Bi-2223, coupling between different planes would be expected to cause the paramagnon to split into three modes. However, the interlayer magnetic exchange interaction, $J_{\perp}$, has been estimated to be well below 10~meV from ARPES measurements of Bi-2212.\cite{Chuang2004} It therefore makes a negligible contribution to the widths observed here. Y-based cuprates spanning the under doped to slightly overdoped region of the phase diagram exhibit paramagnon widths of comparable magnitude $\sim 230$~meV, which are also approximately $\mathbi{Q}$-independent.\cite{LeTacon2011} Data for La$_{2-x}$Sr$_x$CuO$_4$ also show similar widths around the same doping levels.\cite{DeanLSCO2013}

% ELASTIC LINES
Figure~\ref{Fig2} also shows stronger elastic scattering from Bi-2201 than Bi-2223. The strength of the elastic line in RIXS reflects the structural defect density and the surface flatness of the crystals, which varies between different crystals. La$_{2-x}$Sr$_x$CuO$_4$ thin films and cleaved single crystals of Bi-2212 and La$_{2-x}$Ba$_x$CuO$_4$ show similar elastic intensities to the Bi-2223 data;\cite{Braicovich2010, DeanLSCO2013, DeanBSCCO2013, DeanLBCO2013} whereas polished Y-based cuprates display comparable elastic intensities to the Bi-2201 data.\cite{LeTacon2011} Around $(0.2 - 0.3, 0)$ a charge density wave may contribute to the elastic (or quasi-elastic) intensity in Bi-2201 [Ref.~\onlinecite{Comin2014}] although this signal is suppressed in the polarization configuration used here.\cite{DeanLBCO2013} 

% J vs Tc SCALING
We now discuss the implications of Fig.~\ref{Fig3}. The higher energy paramagnons in Bi-2223 compared to Bi-2201 suggest an appealingly simple scenario -- relevant for theories of high-$T_{\rm c}$ superconductivity based on the exchange of magnetic excitations -- that $T_{\rm c}$ is higher in the multilayer cuprates because $J$ is higher.\cite{Scalapino2012} Indeed, $J$ is often the only relevant energy scale in such theories. There are relatively few systematic measurements of $J$ as a function of $n$ that we can compare with the present RIXS results, likely due to difficulties in growing high quality samples. Raman scattering couples to the two magnon density of states at $\mathbi{q} \approx 0$ and this two-magnon peak appears at higher energies in Bi-2212 than Bi-2201,\cite{Sugai2003} consistent with the trend found here.

While higher values of $J$ may play an important role for the increased $T_{\rm c}$ in multilayer cuprates, it should be emphasized that such an effect should only be considered within a single cuprate family. It is clear that this is not the only effect that needs to be invoked to explain the variation of $T_{\rm c}$ between different cuprate families.\cite{*[{See, for example, }] [{}] Ofer2006,*Mallett2013,*Johnston2010}

\begin{figure}
\includegraphics{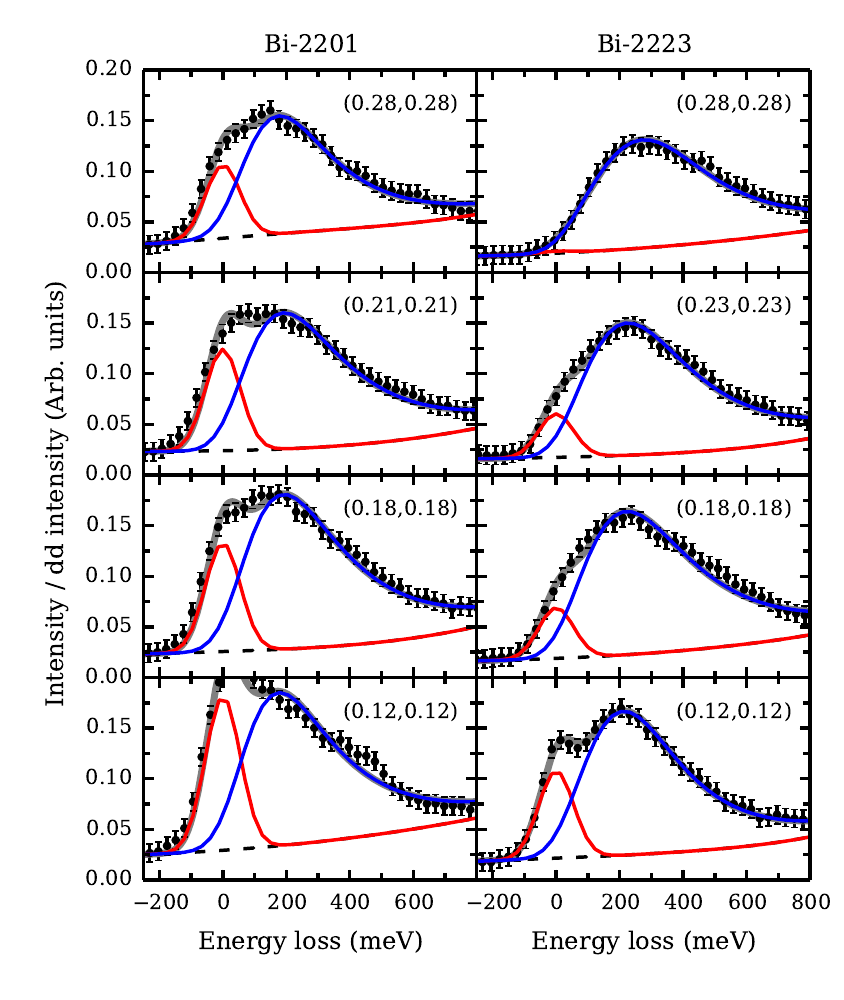} %
\caption{(Color online) Low energy RIXS spectra for Bi-2201 (left) and Bi-2223 (right) along $(\zeta, \zeta)$, where a heavily over-damped, almost non-dispersive excitation is observed. Black points show the data and the solid grey line represents the results of the fitting, which is the sum of a Gaussian elastic scattering contribution (red), an anti-symmetrized Lorentzian (blue), and the smooth background (dashed black line).}
\label{Fig4}
\end{figure}

\begin{figure}
\includegraphics{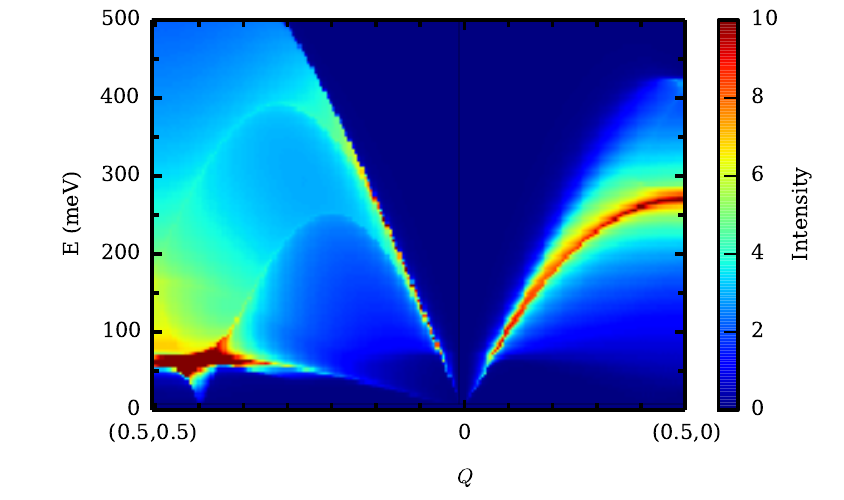} %
\caption{(Color online) Calculations of the imaginary part of the magnetic dynamical susceptibility $\chi^{\prime\prime}(\mathbi{Q},E)$ in the superconducting state based on the renormalized itinerant quasiparticles approach used in Ref.~\onlinecite{Brinckmann2001}. A well-defined paramagnon mode is observed along the antinodal $(0,0)\rightarrow(0.5,0)$ direction whereas the intensity along the nodal $(0,0)\rightarrow(0.5,0.5)$ direction is more diffuse and poorly defined with the exception of the ``resonance mode'' around $(0.5,0.5)$. Apart from the ``resonance mode'' calculations for the non-superconducting state are virtually identical.}
\label{Fig5}
\end{figure}

% PIPI DISCUSSION
We now consider the excitation spectrum along the nodal $(\zeta,\zeta)$ direction, as plotted in Fig.~\ref{Fig4}.  In this direction in reciprocal space, the majority of the spectral weight sits at lower energies relative to the antinodal direction and there is a strong tail of intensity extending out to higher energies. The spectra exhibit very little $\mathbi{Q}$-dependence. This is in distinct contrast to undoped cuprates such as La$_2$CuO$_4$ and Sr$_2$CuO$_2$Cl$_2$ where INS and RIXS spectra show well-defined spin wave excitations in this direction,\cite{Coldea2001, Dean2012, Guarise2010} i.e.\ doping strongly damps the magnetic excitations.  Fitting the same formula to the data in Fig.~\ref{Fig4}, as was done in Fig.~\ref{Fig2}, yields widths that are far larger than the energy of the Lorentzian peak. The leading edge of the lineshape is primarily determined by the energy resolution function and the exact energy of the Lorentzian peak becomes difficult to determine. Indeed, even though the paramagnon lineshape can still be used to describe the data, the lack of dispersion, and the fact that the width of the mode is much larger than the energy, implies very strong scattering and a lack of coherent propagation. This indicates that a paramagnon-based picture is inappropriate in the doped cuprates along the nodal direction. 

% CALCULATIONS
In order to explain this phenomenon, we performed calculations of the imaginary part of the magnetic dynamical susceptibility, $\chi^{\prime\prime}(\mathbi{Q},E)$, based on the renormalized itinerant quasiparticle approach described in Ref.~\onlinecite{Brinckmann2001}. The calculations presented are performed for a single layer cuprate system. Trilayer calculations were also performed. These introduce some additional fine structure, which would not be resolvable with the present energy resolution and the overall form of the results remains the same. The results of the single layer calculation are plotted in Fig.~\ref{Fig5}.\footnote{See supplementary materials at XXXX for a full description.} The prediction captures the well-defined paramagnon feature that is present along the antinodal direction, a feature that arises due to dynamical nesting across the Fermi level. Efficient nesting does not occur for energy transfers above 100~meV along the nodal direction leading to diffuse magnetic weight, and no paramagnon feature, along $(0,0)\rightarrow(0.3,0.3)$. Which qualitatively, although not quantitatively, captures the asymmetry between the nodal and anti-nodal directions. Other itinerant calculations of $\chi^{\prime\prime}(\mathbi{Q},E)$ in the cuprates also show similar phenomenology .\cite{James2012, Zeyher2013, Eremin2013} The fact that itinerant models capture this effect implies an itinerant, or partially itinerant, nature of the cuprate magnetism, at least for the doping, $\mathbi{Q}$ and $E$ scales studied here. 

This result should be considered in the context of the larger debate on the relative merits of localized versus itinerant calculations for describing magnetism in the cuprates.\cite{Vojta2009NatPhys, Vojta2011} ARPES shows the presence of coherent quasiparticles in the electronic structure of optimally doped cuprates,\cite{Damascelli2003} which has motivated itinerant descriptions of the magnetism.\cite{James2012, Zeyher2013, Eremin2013} These should be contrasted with the local-moment magnetism present in the undoped cuprates.\cite{Kastner1998} For example, INS measurements of Bi-2212 have been used to argue that itinerant calculations cannot account for the temperature dependence of the 40~meV magnetic resonance mode around $(0.5, 0.5)$.\cite{Xu2009} This suggests that a combination of local moment and itinerant physics is responsible for the richness of the full spectrum over all $\mathbi{Q}$ and $E$ scales in the doped cuprates. 

Finally, comparing INS and RIXS measurements at finite doping shows that the dispersion of the excitations is not symmetrical about $(0,0)$ and $(0.5,0.5)$. While RIXS shows a strong difference between the nodal $(0,0) \rightarrow (0.25,0.25)$ and antinodal $(0,0) \rightarrow (0.5,0)$ directions, INS demonstrates that around $(0.5,0.5)$ these two different directions both host paramagnon excitations well into the overdoped regime.\cite{Vignolle2007, Lipscombe2007}. This further emphasizes that $(0,0)$ and $(0.5,0.5)$ do not host symmetric dispersions -- such a symmetry is only expected to hold for N\'{e}el ordered cuprates.\footnote{See Ref.~\onlinecite{Dean2014} for a more thorough discussion}

% CONCLUSIONS
To conclude, we present RIXS data on Bi-2201 and Bi-2223. Along the antinodal $(0,0) \rightarrow (0.5,0)$ direction we observe a dispersive damped paramagnon mode with a higher zone boundary energy in Bi-2223 than Bi-2201, which indicates that Bi-2223 hosts stronger effective magnetic exchange interactions than Bi-2201. This fact must be considered, against other proposed explanations, as a possible reason for the enhanced $T_{\rm c}$ in the multilayer cuprates. Along the nodal $(0,0) \rightarrow (0.3,0.3)$ direction we find diffuse, almost non-dispersive spectral weight, in strong contrast to the magnon excitation present in a undoped insulator. We show that calculations of the dynamical magnetic susceptibility based on Ref.~\onlinecite{Brinckmann2001} qualitatively capture the  paramagnon present in the antinodal direction \emph{and} the diffuse magnetic weight in the nodal direction, which indicates that magnetism in the doped cuprates is at least partially itinerant in nature.

\begin{acknowledgments}
We thank Liviu Hozoi, Krzysztof Wohlfeld and Hong Ding for discussions, Paul Olalde-Velasco for support during the experiments and Jason Hancock for sharing his Bi-2223 sample. M.P.M.D.\ and J.P.H.\ are supported by the Center for Emergent Superconductivity, an Energy Frontier Research Center funded by the U.S.\ DOE, Office of Basic Energy Sciences. Work at Brookhaven National Laboratory was supported by the Office of Basic Energy Sciences, Division of Materials Science and Engineering, U.S. Department of Energy under Award No.\ DEAC02-98CH10886. A.J.A.J.'s contribution to this work was supported by the Engineering and Physical Research Council (grant number EP/L010623/1). The experiment was performed at the ADRESS beamline of the Swiss Light Source at the Paul Scherrer Institut.
\end{acknowledgments}

\end{document}